\documentstyle[twocolumn,prb,aps,epsf]{revtex}

\def\upcite#1{$\,$\cite{#1}}

\def\vs{\mbox{\boldmath $s$}}
\def\vn{\mbox{\boldmath $n$}}

\def\vz{\mbox{\boldmath $z$}}
\def\vr{\mbox{\boldmath $r$}}

\def\vv{\mbox{\boldmath $v$}}

\begin{document}
\draft
\preprint{}
\wideabs{
%%%%%%%%%%%%%%%%%%%%
\title{Dynamics of vortex tangle without mutual friction
in superfluid $^4$He}
\author{Makoto Tsubota$^1$, Tsunehiko Araki$^1$,
and Sergey K. Nemirovskii$^2$}
\address{$^1$Department of Physics,
Osaka City University, Sumiyoshi-Ku,
Osaka 558-8585, Japan \\
$^2$Institute of Thermophysics, Academy of Science, Novosibirsk
630090, Russia}
\date{Received
\hspace{25mm}
}
\maketitle
\begin{abstract}
A recent experiment has shown that a tangle of quantized vortices in
superfluid $^4$He decayed even at mK temperatures where the normal fluid
was negligible and no mutual friction worked.
Motivated by this experiment, this work studies numerically the dynamics
of the vortex tangle without the mutual friction, thus showing that
a self-similar cascade process, whereby large vortex loops break up to
smaller ones, proceeds in the vortex tangle and is closely related
with its free decay.
This cascade process which may be covered with the mutual friction at
higher temperatures is just the one at zero temperature
Feynman proposed long ago.
The full Biot-Savart calculation is made for dilute vortices, while the
localized induction approximation is used for a dense tangle.
The former finds the elementary scenario: the reconnection of the vortices
excites vortex waves along them and makes them kinked, which could be
suppressed if the mutual friction worked.
The kinked parts reconnect with the vortex they belong to, dividing into
small loops.
The latter simulation under the localized induction approximation shows
that such cascade process actually proceeds self-similarly in a dense tangle
 and continues to make small vortices.
Considering that the vortices of the interatomic size no longer keep the
picture of vortex, the cascade process leads to the decay of the vortex
line density.
The presence of the cascade process is supported also by investigating
the classification of the reconnection type and the size distribution of
vortices.
The decay of the vortex line density is consistent with the solution
of the Vinen's equation which was originally derived on the basis of
the idea of homogeneous turbulence with the cascade process.
The cascade process revealed by this work is an intrinsic process in the
superfluid system free from the normal fluid.
The obtained result is compared with the recent Vinen's theory which
discusses the Kelvin wave cascade with sound radiation.

\end{abstract}
\pacs{67.40.Vs, 67.40.Bz}
}
%
%%%%%%%%%%%%%%%%%%%%
\section{INTRODUCTION}
Superfluid $^4$He (Helium II) behaves like an irrotational ideal fluid,
whose characteristic phenomena can be explained well by the Landau two-fluid
model.
However, superflow becomes dissipative (superfluid turbulence) above some
critical velocity.
The concept of superfluid turbulence was introduced by Feynman\upcite{Feynman}
who stated that the superfluid turbulent state consists of a disordered set
of quantized vortices,\upcite{Donnelly,Nemi} called vortex tangle(VT).
Reminding the inertial range of the classical-fluid turbulence, Feynman
proposed that VT undergoes the following cascade process.
At zero temperature, a large distorted vortex loop breaks up to smaller
loops through reconnections, and the cascade process continues self-similarly
down to the order of the interatomic scale.
At finite temperatures, however, normal fluid collides with vortices and
takes energy from them.

This idea was developed further by Vinen.
In order to describe an amplification of a temperature difference at the
ends of a capillary retaining thermal counterflow, Gorter and Mellink
introduced some additional interactions between the normal fluid and
superfluid.\upcite{Gorter}
Through experimental studies of the second-sound attenuation, Vinen
considered this Gorter-Mellink mutual friction in relation to the
macroscopic dynamics of the VT.\upcite{Vinen}
Assuming homogeneous superfluid turbulence, Vinen obtained an evolution
equation for the vortex line density (VLD) $L(t)$, what we call the Vinen's
equation
\begin{equation}
\frac{dL}{dt}=\alpha |\vv_{\rm ns}|L^{3/2}-\chi_2 \frac{\kappa}{2\pi}L^2,
\label{Vinen}
\end{equation}
where $\alpha$ and $\chi_2$ are parameters dependent on temperature and
$\vv_{\rm ns}$ is the relative velocity between the normal flow and
superflow, $\kappa$ the quantized circulation.
This Vinen's theory could describe well a large number of observations
of mostly stationary cases.

However the nonlinear and nonlocal dynamics of vortices had long delayed
the progress in further microscopic understanding of the VT.
It was Schwarz that broke through.\upcite{Schwarz85,Schwarz88}
His most important contribution was that the direct numerical simulation
of vortex dynamics connected with the scaling analysis enabled us
to calculate such physical quantities as the VLD, some anisotropic
parameters, the mutual friction force, etc.
The observable quantities obtained by Schwarz's theory agree well with
the experimental results of the steady state of the VT.
This research field pioneered by Schwarz has revealed many problems of
vortex dynamics, such as the flow properties in channels
\upcite{Samuels92,Barenghi,Aarts,Penz}, sideband instability
of Kelvin waves,\upcite{Samuels90} vortex array in rotating
superfluid,\upcite{Tsubota95} vortex pinning.\upcite{Tsubota93,Tsubota94}

The mutual friction plays an important role in the above vortex dynamics.
The stationary state of the VT Schwarz obtained is self-sustaining, and
realized by the competition between the excitation and dissipation due to the
mutual friction subject to the $\vv_{\rm ns}$ field, as described in the
next section.
Hence the system free from the mutual friction cannot sustain the
stationary VT.

Compared with the steady state, there have been less studies of the
transient behavior of the VT.
Although the transient behavior generally refers to both the growth and
decay process, this paper considers only the decay of the VT after the
driving velocity is suddenly reduced to zero.
The early measurements by Vinen\upcite{Vinen} and the later
ones\upcite{Milliken,Rozen} observed a decay of the VT which was
consistent with the Vinen's equation (\ref{Vinen}) with only the decay
term, although Schwarz and Rozen\upcite{Rozen} coupled the Vinen's
equation with the hydrodynamical equations of the normal flow and
the superflow in order to explain a slow decay following an initial
rapid decay they observed.
Apart from these experiments on thermal counterflow, the decay of
vorticity in turbulence generated by towing a grid was studied
recently.\upcite{Smith,Stalp}
This turbulence is expected to be homogeneous and isotropic.
The experimental results may be understood by the picture that
the mutual friction can be so strong that the normal fluid and the
superfluid lock together, behaving effectively like a single
fluid.\upcite{BarenghiSamuels,Samuels99}
The experimental results are compared with the change in the turbulent
energy spectrum which includes the Kolmogorov law.

Both these numerical and experimental results are much affected by the
mutual friction.
However, recently, Davis et al.\upcite{Davis} observed that vortices did
decay even
at mK temperatures where the normal fluid density became vanishingly
small and, as a consequence, the mutual friction did not work effectively.
The vortices were created by a vibrating grid, and detected by their trapping
of negative ions.
The first important point is that the vortices actually decay at
such low temperatures.
The second is that the decay rate becomes independent of temperature
below $T\simeq 70$mK.
It is unclear how the vortices decay.
This experimental work, which is just preliminary at present, can develop
a new research field of superfluid or vortex dynamics at mK temperatures;
it can reveal some essence that may be covered with the normal fluid at
higher temperatures.

Motivated by this experimental work, we study numerically the vortex dynamics
without the mutual friction.
The calculation under the localized induction approximation(LIA) is made
for the dense VT, while the full Biot-Savart calculation for the more dilute
vortices.
The absence of the mutual friction makes the vortices kinked, which
promotes vortex reconnections.
Consequently small vortices are cut off from a large one through
the reconnections.
The resulting vortices also follow the self-similar process to break
up to smaller ones.
Although our formulation cannot describe the final destiny of the
minimum vortex, the decay of the VT is found to be connected with
this cascade process, which is just the cascade process at zero
temperature Feynman proposed.\upcite{Feynman}

The contents of this paper are the following.
Section II describes the equations of motion of vortices and the
method of numerical calculation.
Section III studies the dynamics of dilute vortices under the full
Biot-Savart law both without and with solid boundaries;
this calculation reveals the essence of the cascade process.
The dynamics of the dense VT under the LIA is discussed in Sec. IV.
The obtained results are compared with the solution of the Vinen's
equation in Sec. V.
The agreement is good, which supports the picture of the cascade process.
The decay of the VT subject to the mutual friction is discussed too.
Section VI is devoted to conclusions and discussions.

%
%%%%%%%%%%%%%%%%%%%%
\section{EQUATIONS OF MOTION AND NUMERICAL SIMULATION}

A quantized vortex is represented by a filament passing through the fluid
and has a definite direction corresponding to its vorticity.
Except for the thin core region, the superflow velocity field has a
classically well-defined meaning and can be described by ideal fluid
dynamics.
The velocity produced at a point $\vr$ by a filament is given by the
Biot-Savart expression :
\begin{equation}
\vv_{s,\omega}=\frac{\kappa}{4\pi}\int_{\cal L} \frac{(\vs_1 - \vr)
\times d\vs_1}{|\vs_1-\vr|^3},
\label{BS}
\end{equation}
where $\kappa$ is the quantized circulation.
The filament is represented by the parametric form $\vs = \vs(\xi, t)$,
$\vs_1$ refers to a point on the filament and the integration is taken
along the filament.
The Helmholtz's theorem for a perfect fluid states that the vortex moves
with the superfluid velocity at the point.
Attempting to calculate the velocity $\vv_{s,\omega}$ at a point $\vr=\vs$
on the filament makes the integral diverge as $\vs_1 \rightarrow \vs$.
To avoid it, we divide the velocity $\dot{\vs}$ of the filament at the point
$\vs$ into two components\upcite{Schwarz85}:
\begin{equation}
\dot{\vs} =\frac{\kappa}{4\pi}\vs' \times \vs'' \ln \left(
\frac{2(\ell_+ \ell_-)^{1/2}}
{e^{1/4} a_0}\right) + \frac{\kappa}{4\pi}\int_{\cal L}^{'} \frac{(\vs_1 -
\vr)
\times d\vs_1}{|\vs_1-\vr|^3}.
\label{sdot}
\end{equation}
The first term shows the localized induction field arising from a curved
line element acting on itself, and $\ell_+$ and $\ell_-$ are the lengths
of the two adjacent line elements that hold the point $\vs$ between, and
the prime denotes differentiation with respect to the arc length $\xi$.
The mutual perpendicular vectors $\vs'$, $\vs''$ and $\vs' \times \vs''$
point along the tangent, the principal normal and the binormal at the point
$\vs$, respectively,
and their magnitudes are 1, $R^{-1}$ and $R^{-1}$, where $R$ is the local
radius of curvature.
The parameter $a_0$ is a cutoff parameter corresponding to a core radius.
Thus the first term tends to move the local point $\vs$ with a velocity
inversely proportional to $R$, along the binormal direction.
The second term represents the nonlocal field obtained by carrying
out the integral of Eq. (\ref{BS}) along the rest of the filament.
The approximation that describes the vortex dynamics neglecting the nonlocal
terms and replacing Eq. (\ref{sdot}) by
\begin{equation}
\dot{\vs} = \beta \vs' \times \vs''
\end{equation}
is called the localized induction approximation(LIA).
Here the coefficient $\beta$ is defined by
\begin{equation}
\beta = \frac{\kappa}{4\pi} \ln \left( \frac{c<R>}{a_0}\right),
\label{beta}
\end{equation}
where $c$  is a constant of order 1 and $(\ell_+ \ell_-)^{1/2}$ is replaced
by the characteristic radius $<R>$.

When boundaries are present, the boundary-induced field $\vv_{s,b}$ is added
to $\vv_{s,\omega}$ so that the boundary condition $(\vv_{s,\omega}+\vv_{s,b})
\cdot \hat{\vn} =0$ can be satisfied.
If the boundaries are specular plane surfaces, $\vv_{s,b}$ is just the field
by an image vortex made by reflecting the vortex into the plane and reversing
its direction of the vorticity.
Some other applied field $\vv_{s,a}$, if present, is added, which results in
the total velocity $\dot{\vs}_0$ of the vortex filament without dissipation:
\begin{eqnarray}
\dot{\vs}_0 =&&\frac{\kappa}{4\pi}\vs' \times \vs'' \ln \left(
\frac{2(\ell_+ \ell_-)^{1/2}}{e^{1/4} a_0}\right) \nonumber\\
&+& \frac{\kappa}{4\pi}\int_{\cal L}^{'} \frac{(\vs_1 - \vr)
\times d\vs_1}{|\vs_1-\vr|^3} +\vv_{s,b}(\vs)+\vv_{s,a}(\vs).
\label{s0dot}
\end{eqnarray}
\noindent
At finite temperatures the mutual friction due to the interaction
between the vortex core and the normal fluid flow $\vv_n$ is taken into
account.
Then the velocity of a point $\vs$ is given by\upcite{Donnelly}
\begin{equation}
\dot{\vs} =\dot{\vs}_0 + \alpha \vs' \times (\vv_n - \dot{\vs}_0)
- \alpha' \vs' \times [\vs' \times (\vv_n - \dot{\vs}_0)],
\label{sdotmf}
\end{equation}
where $\alpha$ and $\alpha'$ are the temperature-dependent friction
coefficients, and $\dot{\vs}_0$ is calculated from Eq.(\ref{s0dot}).
All calculations in this work are made for
$\alpha' =0$.\upcite{Schwarz85}

As discussed by Barenghi and Samuels\upcite{BarenghiSamuels},
this formulation is essentially kinematic in the sense that the
driving flows $\vv_n$ and $\vv_{s,a}$ are constant, that is, they
only act on the vortex dynamics but are never affected by it.
When the dynamics of the driving flows is concerned, it should be
coupled selfconsistently to the vortex dynamics.
However, since this work studies the system without the
normal fluid and the driving superflow, this formulation will be
useful to describe correctly the vortex dynamics, except for
the phenomena that is concerned with the vortex core region, such as
vortex reconnection, nucleation and annihilation.

Studying the vortex dynamics without the mutual friction needs to
understand qualitatively the role of the mutual
friction.\upcite{Schwarz85}
Let us assume the LIA and neglect the term with $\alpha'$.
Then Eqs. (\ref{s0dot}) and (\ref{sdotmf}) are reduced to
\begin{equation}
\dot{\vs}=\beta \vs' \times \vs'' + \vv_{s,a}+ \alpha \vs' \times
(\vv_n-\vv_{s,a}-\beta \vs' \times \vs'').
\label{scaling}
\end{equation}
If the mutual friction is absent, the dynamics due to only the
self-induced velocity conserves the total line length of vortices.
Under the above mutual friction, one can easily find that when the
applied relative flow $\vv_n-\vv_{s,a}$ blows against the local
self-induced velocity $\beta \vs' \times \vs''$, the mutual friction
always shrinks the vortex line locally.
On the other hand, the relative flow along the self-induced velocity yields
a critical radius of curvature
\begin{equation}
R_c \simeq \frac{\beta}{|\vv_n-\vv_{s,a}|}.
\end{equation}
When the local radius $R$ at a point on a vortex is smaller than $R_c$,
the vortex will shrink locally, while the vortex of $R > R_c$ balloons out.
Thus it should be noted that the mutual friction plays the dual role
of the growth and decay of vortex line length.
This dual role of the mutual friction sustains the steady state of the
VT subject to the applied flow, where the highly curved structure whose
local radius of curvature is less than $R_c$ will be smoothed out.
If this applied field is absent, $R_c$ becomes infinite so that
an arbitrary curved configuration of vortex lines shrinks away.

Here we will describe shortly the dynamical scaling discussed by
Swanson and Donnelly,\upcite{Swanson} and Schwarz\upcite{Schwarz88},
which is necessary for understanding the cascade process of the
VT dynamics.
Using the LIA and absorbing the factor $\beta$ into reduced time
$t_0=\beta t$ and velocity $\vv_0=\vv/\beta$, Eq. (\ref{sdotmf})
becomes
\begin{eqnarray}
\frac{\partial \vs}{\partial t_0} =&&\vs' \times \vs''+ \vv_{s,0}
+ \alpha \vs' \times (\vv_{n,0}-\vv_{s,0} - \vs' \times \vs'') \nonumber\\
&-& \alpha' \vs' \times [\vs' \times (\vv_{n,0}-\vv_{s,0}
- \vs' \times \vs'')].
\end{eqnarray}
This equation is invariant under the scale transformation:
\begin{eqnarray}
\vs=\lambda \vs^*,\quad \xi=\lambda \xi^*,
\quad t_0=\lambda^2 t_0^*, \nonumber \\
\vv_{n,0}=\lambda^{-1} \vv_{n,0}^*, \quad \vv_{s,0}=\lambda^{-1} \vv_{s,0}^*.
\end{eqnarray}
Accordingly, if all space coordinates of a system are reduced by a factor
$\lambda(<1)$, the dynamics of the new system will look like the same
as that of the old one, except that the velocity increases by $\lambda^{-1}$
and the time passes more rapidly by $\lambda^2$.
In other words, a small vortex loop whose configuration is similar to a
large one but size is reduced by $\lambda$ follows the similar motion
whose time scale shortens by $\lambda^2$ compared with the large one.

Some important quantities which are useful for characterizing the VT
will be introduced.\upcite{Schwarz88}
The vortex line density(VLD) is
\begin{equation}
L= \frac{1}{\Omega} \int_{\cal L} d \xi,
\end{equation}
where the integral is made along all vortices in the sample
volume $\Omega$.
Even though the VT may be homogeneous, it need not generally isotropic.
The anisotropy of the VT which is made under the counterflow $\vv_{ns}$
is represented by the dimensionless parameters
\begin{mathletters}
\begin{eqnarray}
I_{\|}=\frac{1}{\Omega L} \int_{\cal L} [1 - (\vs' \cdot \hat{\vr}
_{\|})^2 ]d \xi, \\
I_{\bot}=\frac{1}{\Omega L} \int_{\cal L} [1 - (\vs' \cdot \hat{\vr}
_{\bot})^2 ]d \xi, \\
I_{\ell} \hat{\vr}_{\|} =\frac{1}{\Omega L^{3/2}} \int_{\cal L}
\vs' \times \vs" d \xi.
\end{eqnarray}
\end{mathletters}
Here $\hat{\vr}_{\|}$ and $\hat{\vr}_{\bot}$ stand for unit vectors
parallel and perpendicular to the $\vv_{ns}$ direction.
The symmetry generally yields the relation $I_{\|}/2+I_{\bot}=1$.
If the VT is isotropic, the average of these measures are
$\bar{I}_{\|}=\bar{I}_{\bot}=2/3$ and $\bar{I}_{\ell} =0$.

The method of the numerical calculations is similar to that of Schwarz
\upcite{Schwarz85} and described in our previous paper\upcite{Tsubota93}.
A vortex filament is represented by a single string of points.
The vortices configuration of a moment determines the velocity field
in the fluid, thus moving the points on vortex filaments by Eqs.
(\ref{s0dot}) and (\ref{sdotmf}).
Both local and nonlocal terms are represented by means of line elements
connecting two adjacent points.
As discussed in Ref. \onlinecite{Schwarz85}, the explicit forward integration
of the local term may be numerically unstable.
To prevent the difficulty, a modified hopscotch algorithm is adopted.
As the vortex configuration develops and, particularly, two vortices
approach each other, the length of a line element can change.
Then it is necessary to add or remove points properly so that
the local resolution does not lose(an adaptive meshing routine).
Through the cascade process described in Sec. III, a large vortex
can break up many times, eventually to a small one whose size
is less than the space resolution, i.e., the distance
between neighboring points on the filament.
Of course the numerical calculation generally cannot follow the
dynamics beyond its space resolution.
Thus such vortices are eliminated numerically; the physical
justification of this cut-off procedure will be discussed in Sec. IV.

How to deal with vortex reconnection is very important in the
simulation of the VT.
The numerical study of the incompressible Navie-Stokes fluid showed
that the close interaction of two vortices leads to their reconnection,
chiefly because of the viscous diffusion of the vorticity.\upcite{Boratav}
Koplik and Levine solved directly the Gross-Pitaevskii equation to show
the two close quantized vortices reconnect even in a inviscid
fluid.\upcite{Koplik}
Of course our numerical method for vortex filaments cannot represent
the reconnection process itself.
However Schwarz\upcite{Schwarz85} and the authors\upcite{Tsubota92} simulated
the vortex dynamics near the reconnection using the full Biot-Savart law.
When two vortices approach each other, let us define a critical
distance\upcite{Schwarz85}
\begin{equation}
\Delta \simeq 2R /\ln (c<R>/a_0),
\end{equation}
at which the nonlocal field from the other becomes comparable to its own
local-induced field.
Two vortices approaching within $\Delta$ cause local twists on each other
so that they become antiparallel at the closest place, even though they
are not antiparallel initially.
Then local cusps connecting these two develop,
which will lead to reconnection.
After the reconnection, two vortices run away rapidly from each other
owing to their self-induced velocity.
Considering both the full Biot-Savart calculation and the results
of Ref.(\onlinecite{Koplik}), it will be reasonable to assume that
two close filaments would reconnect.
This assumption has an important meaning beyond a numerical expedient.
The numerical simulation of the dense VT forces us to use the LIA,
because the full Biot-Savart calculation requires much computing time.
The LIA is expected to be a good approximation (to order 10$\%$)
provided the inter-vortex spacing is enough large.
However, when two vortices approach each other more closely than
$\Delta$, the nonlocal field becomes not negligible in reality.
All the effects coming from such nonlocal field may be thought to be
 renormalized artificially by making the vortices reconnect.
In the numerical simulation of the VT, Schwarz assumed that vortices
which pass within $\Delta$ are reconnected with unit probability.
He noticed that the details of when and how the vortices are reconnected
have no significant influence on the behavior of the VT, while
the judgment by this $\Delta$ can make unphysical reconnections.
For example, two almost straight vortices must reconnect even if
they are very apart, because their large radius $R$ of curvature
results in the large $\Delta$.
The full Biot-Savart calculation\upcite{Tsubota92} shows that
two vortices that once approach within $\Delta$ can get away without
reconnecting.
Hence, in contrast to the method of Schwarz, this work reconnects
the vortices which pass within not $\Delta$ but the space resolution
$\Delta \xi$, for both the LIA and the full biot-Savart calculations.
The concrete procedure is the following.
Every vortex initially consists of a string of points at regular
intervals of $\Delta \xi$.
The subsequent vortex motion can change the intervals of two adjacent
points, yet the above adaptive meshing routine keeps each interval
almost $\Delta \xi$.
When a point on a vortex approaches another point on another vortex
more closely than the fixed space resolution $\Delta \xi$, we join
these two points and reconnect the vortices.
Before and after the reconnection, the local line length may increase
or decrease by a small quantity less than $\Delta \xi$.
This procedure is best for the filament reconnection under the
full Biot-Savart calculation.
The dependence of the LIA dynamics on $\Delta \xi$ will be discussed
in Sec. IV.

The numerical space resolution $\Delta \xi$ and the time resolution
$\Delta t$ will be described for each calculation.
For example, the dense tangle in a 1cm$^3$ cube shown in
Fig. \ \ref{decay of VT} (a) is calculated using $\Delta \xi=
1.83 \times 10^{-2}$cm, $\Delta t = 1.0 \times 10^{-3}$sec., $N \simeq
16,000$points.
Then, as described in Sec. IV, the VLD is conserved properly under
the LIA, except for at each moment of reconnection.

%
%%%%%%%%%%%%%%%%%%
\section{DECAY OF DILUTE VORTICES}
This section will investigate the dynamics of dilute vortices by the
full Biot-Savart law described by Eq. (\ref{s0dot}).

\begin{figure}
\epsfxsize=8cm \epsfbox{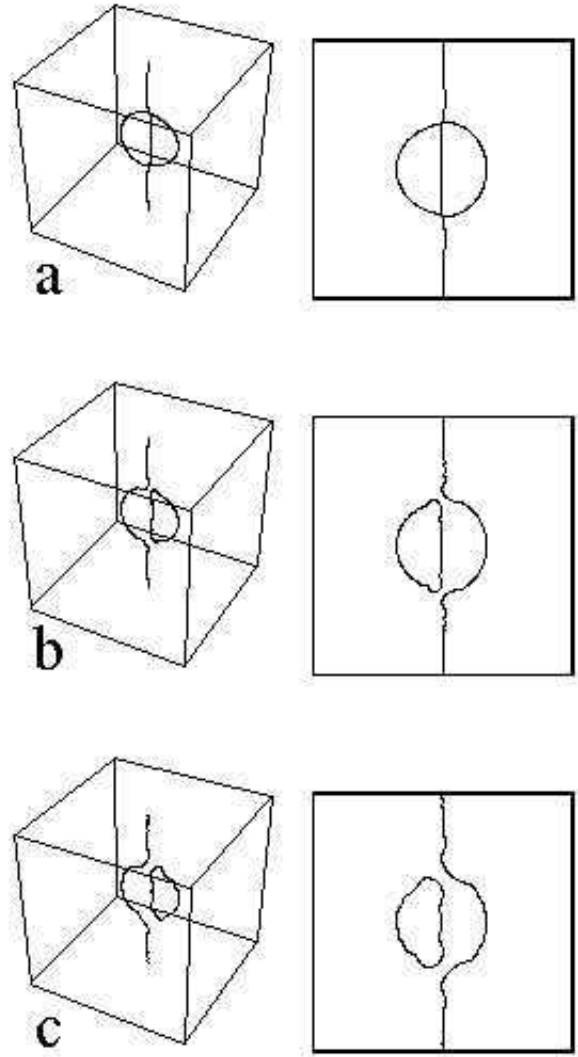}
\vspace*{3mm}
\caption{Collision of a straight vortex and a ring by the full Biot-Savart
calculation without the mutual friction. The right column shows the side view
of the left. }
\label{nonlocal0K}
\end{figure}

\begin{figure}
\epsfxsize=8cm \epsfbox{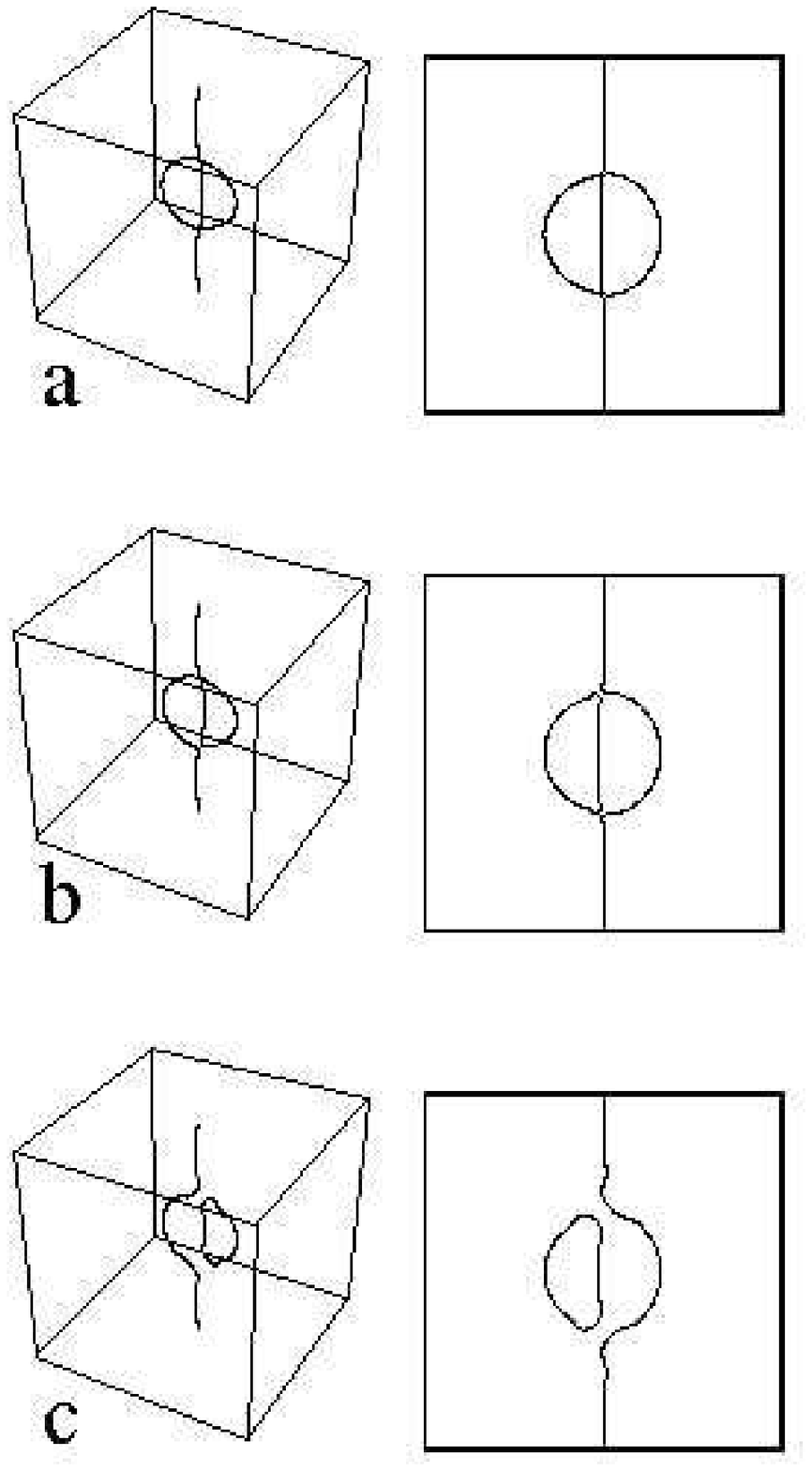}
\vspace*{3mm}
\caption{Collision of a straight vortex and a ring by the full Biot-Savart
calculation with the mutual friction($\alpha=0.1$). The right column shows
the side view of the left. }
\label{nonlocal1.6K}
\end{figure}

We will begin with the collision of a straight vortex line and a moving
ring in order to investigate what happens after the reconnection.
Figure \ \ref{nonlocal0K} shows the motion without the mutual friction.
Toward the reconnection, the ring and the line twist themselves so
that they become locally antiparallel at the closest place(Fig.
\ \ref{nonlocal0K} (a)).
After the reconnection(Fig. \ \ref{nonlocal0K} (b) and (c)), the
resulting local cusps\upcite{Schwarz85,Tsubota92} propagate along the
vortices, exciting vortex waves.
As shown in Fig. \ \ref{nonlocal1.6K}, the dynamics with the mutual
friction($\alpha = 0.1$) is similar, but there is a noticeable
difference; the vortices are relatively smooth because of that
smoothing effect of the mutual friction.
For comparison, we calculated the dynamics under the LIA without the mutual
friction.
Although the twist due to the nonlocal interaction is absent,
the behavior is similar to that of Fig. \ \ref{nonlocal0K}.
It should be noted that the total line length under the LIA without the
mutual friction
is properly conserved within the numerical error except for at the moment
of reconnection, while it is just lengthened by the nonlocal
interaction in Fig. \ \ref{nonlocal0K}.

\begin{figure}
\epsfxsize=8cm \epsfbox{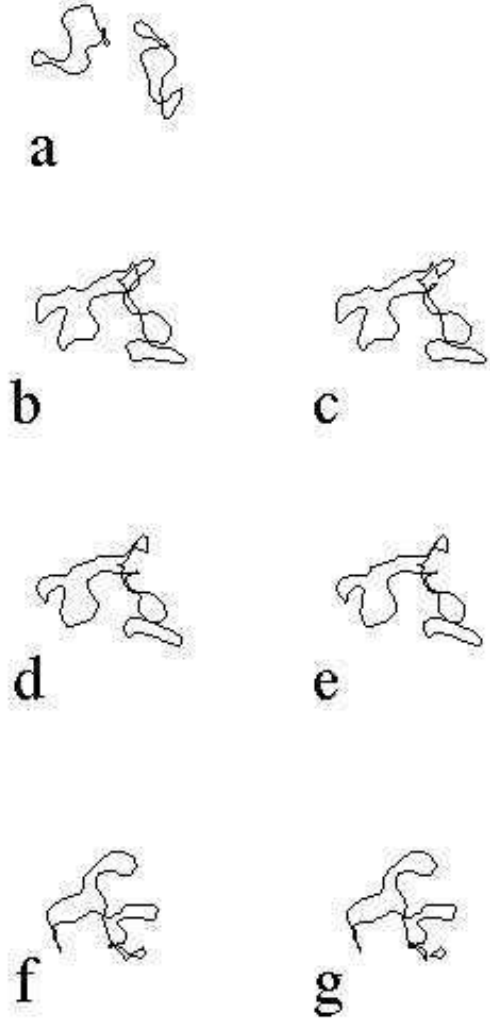}
\vspace*{3mm}
\caption{Typical motion of two vortices by the full
Biot-Savart calculation. They approach(a), and reconnect(b) to
be combined to one loop(c). Afterthat it is kinked(d) to cut off
a small loop from itself(e). The same process occurs again((f)
and (g)).}
\label{scenario}
\end{figure}

\begin{figure}
\epsfxsize=8cm \epsfbox{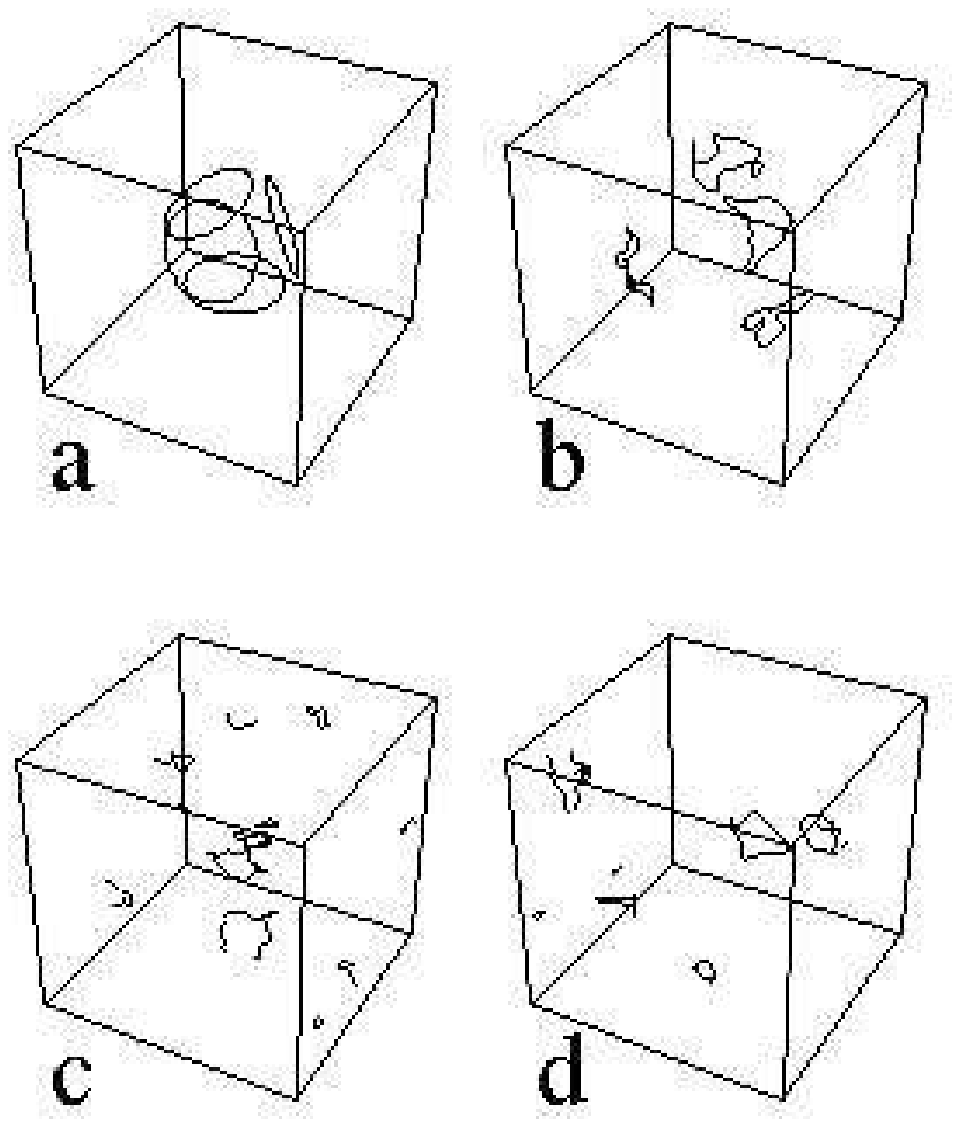}
\vspace*{3mm}
\caption{Motion of four vortex rings by the full Biot-Savart calculation
without the mutual friction. The system is a 1cm$^3$ cube and the periodic
boundary conditions are used in all directions. The time is $t=0$s(a),
30s(b), 150s(c) and 500s(d). }
\label{dilute}
\end{figure}

A typical scenario that vortex loop follows is shown in Fig.
\ \ref{scenario}, which is a part of the process of Fig. \ \ref{dilute}.
Two vortex loops approach each other to reconnect, thus becoming
one loop.
The reconnection excites vortex waves along the loop and makes it
kinked.
The kinked parts reconnect with the loop itself they belong to,
thereby dividing into smaller loops.
Then we are afraid that these kinks may arise from bad numerical methods,
which can be denied by the following reasons.
First, the calculation is made by enough mesh points even when there
appear kinks.
For example, even the left vortex in  Fig. \ \ref{scenario} (a) is
represented by about 60 points.
Secondly, as described in the last paragraph, we confirm that the
total line length is conserved in the dynamics under the LIA without
the mutual friction.
Thirdly, a circular vortex ring is found to move at the expected speed
without making kinks, which was proposed by Schwarz\upcite{Schwarz85}
as a method that checks the numerical scheme.

Considering the above results, we will study the dynamics of dilute
vortices with and without the mutual friction.
The computation sample is taken to be a cube of size 1cm.
The calculation is made by the space resolution $\Delta \xi=1.83 \times
 10^{-2}$cm and the time resolution $\Delta t = 1.0 \times 10^{-3}$sec.
The initial configuration consists of four identical vortex rings
placed symmetrically.
We will study first the system subject to the periodic boundary
conditions in all directions, that is, any vortex
leaving the volume appears to reenter it from the opposite face, and
next that surrounded by smooth, rigid walls.

\begin{figure}
\epsfxsize=8cm \epsfbox{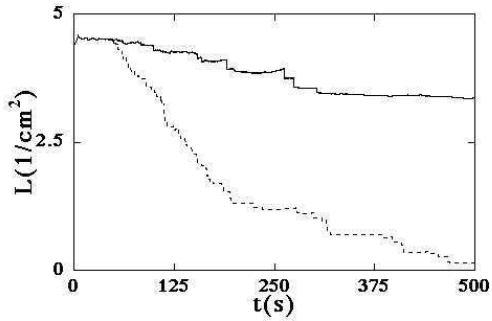}
\vspace*{3mm}
\caption{Decay of the VLD $L(t)$. The solid and dotted lines refer to
the dynamics of Fig.\ \ref{dilute} subject to the periodic boundary
conditions and that of Fig. \ \ref{dilute wall}
confined by solid walls, respectively.}
\label{dilute L}
\end{figure}

\begin{figure}
\epsfxsize=8cm \epsfbox{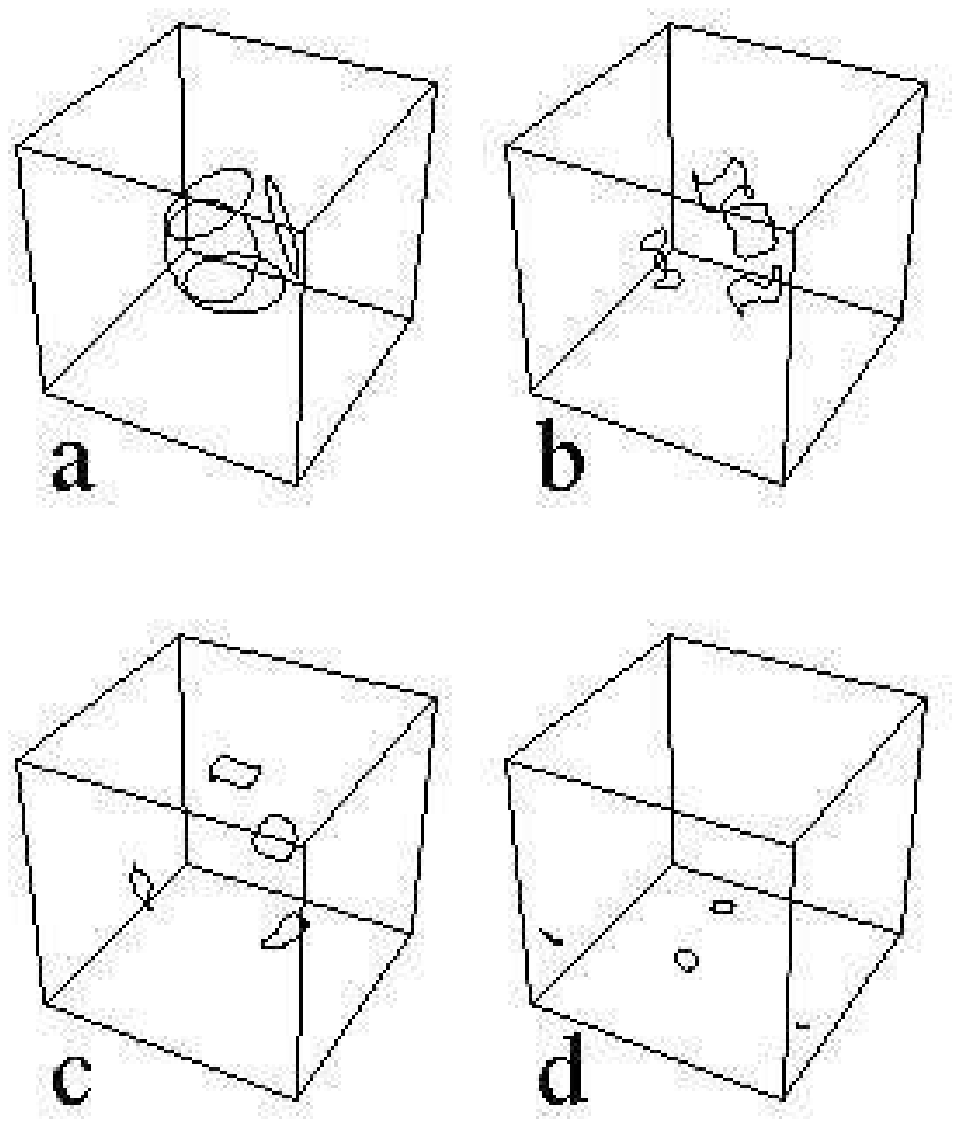}
\vspace*{3mm}
\caption{Motion of four vortex rings by the full Biot-Savart calculation
with the mutual friction($\alpha=0.1$). The system is a 1cm$^3$ cube and
the periodic boundary conditions are used in all directions. The time
is $t=0$s(a), 10s(b), 20s(c) and 40s(d).}
\label{dilute mf}
\end{figure}

Figure \ \ref{dilute} shows the dynamics in the absence of the mutual
friction.
Four rings move toward the center of the cube by their self-induced
velocity to make the first reconnection(a); the four rings resulting
afterthat move outside oppositely(b).
During the motion, they become kinked because of that mechanism
described previously, and cut off their small kinked parts by
reconnection.
The periodic boundary conditions make the vortices collide
repeatedly((c) and (d)), so that this self-similar process continues
down to the scale of the space resolution below which the vortices
are supposed to be eliminated numerically.
This can be considered as the degenerate cascade process that follows
the cascade decay process of the dense tangle investigated
in the next section.
Figure \ \ref{dilute L} shows the decay of the VLD $L(t)$ in the
process of Fig. \ \ref{dilute}.
When two vortices approach each other, the nonlocal interaction can
stretch them, which sometimes causes just a little increase in $L(t)$.
However the superior cascade process decreases the VLD
as a whole.
The effect of the mutual friction is shown in Fig. \ \ref{dilute mf}.
The difference is apparent.
The mutual friction smoothes and shrinks the vortex lines before lots
of reconnection.

\begin{figure}
\epsfxsize=8cm \epsfbox{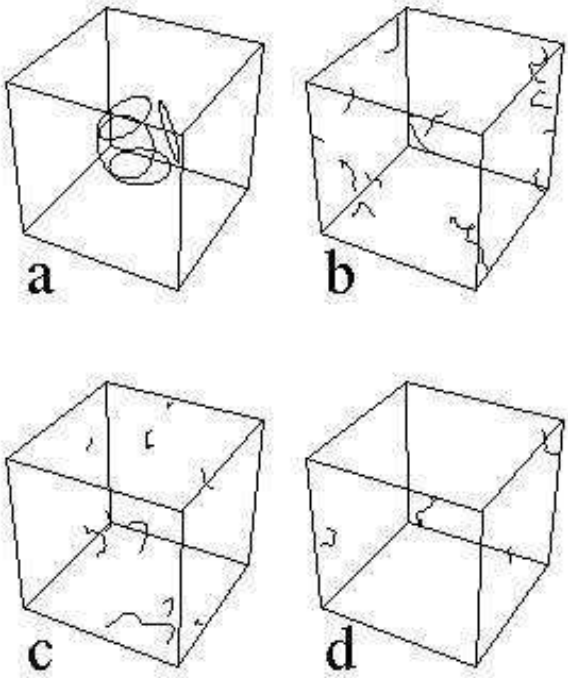}
\vspace*{3mm}
\caption{Motion of four vortex rings by the full Biot-Savart calculation
without the mutual friction. The system is a 1cm$^3$ cube and the system
is confined by solid walls. The time is $t=0$s(a),
100s(b), 150s(c) and 300s(d). }
\label{dilute wall}
\end{figure}

Figure \ \ref{dilute wall} shows the dynamics with boundaries,
starting from the same initial conditions.
Although the early behavior (a) is similar to that of Fig. \ \ref{dilute},
all vortices collide with the boundaries and get attached there (b),
afterthat behaving differently.
Running along the walls (c) and colliding with the faces of the cube,
they become kinked and broken up through
the cascade process, ending in a degenerate state (d).
As shown in Fig. \ \ref{dilute L}, the VLD with the boundaries
decays faster than that without boundaries.
Under the periodic boundary conditions, the vortices collide
only when they happen to meet each other in the volume.
In the presence of solid boundaries, however, the vortex which
runs along one boundary surface of the cube collides with its
image vortex whenever it comes across another face.
Thus the presence of the boundaries causes more reconnections and
promotes the cascade process, which reduces VLD faster than
the case of periodic boundary condition.
We find that the system whose size of the cube is enlarged by
a factor delays the decay of the VLD by the same factor,
which supports strongly this scenario.

%
%%%%%%%%%%%%%%%%%%%%
\section{DECAY OF THE VORTEX TANGLE}

This section studies the free decay of the dense VT without mutual friction
under the LIA.
The decay of dilute vortices described in the last section follows
this decay of the VT.

Throughout this section, the computation sample is taken to be a cube
of size 1cm.
The calculation is made by the space resolution $\Delta \xi=1.83 \times
 10^{-2}$cm and the time resolution $\Delta t = 1.0 \times 10^{-3}$sec.
The one set of faces is subject to periodic boundary conditions.
The other two sets of faces are treated as smooth, rigid boundaries,
in which case vortices approaching the faces reconnect to them and
their ends can move smoothly along the wall.
The reason why we do not adopt the periodic boundary conditions in all
directions is that then an artificial mixing process is necessary
for obtaining an isotropic VT.\upcite{Schwarz88}

\begin{figure}
\epsfxsize=8cm \epsfbox{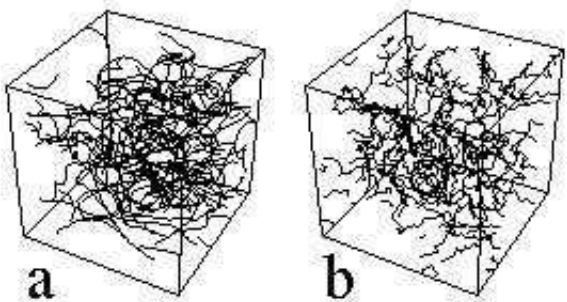}
\vspace*{3mm}
\caption{Example of a VT with (a) and without (b) the mutual friction.}
\label{comp. VT}
\end{figure}

How to prepare the initial VT for free decay
 follows the method used by Schwarz.\upcite{Schwarz88}
An initial state of six vortex rings is allowed to develop under
a pure driving normal flow $\vv_n = v_n \hat{\vz}$, where $\hat{\vz}$ is
parallel to the direction along which the periodic boundary condition
is used.
This process should be made through the dynamics with the mutual
friction($\alpha = 0.1$), because the vortices free from the mutual friction
never
grow to a tangle as shown by Eq. (\ref{scaling}).
Although Schwarz continued the calculation until the vortices
grew up to a steady self-sustaining state, we will take a
growing VT at a moment to prepare a initial state for the
simulation of the free decay.
Figure \ \ref{comp. VT}(a) shows a example of the transient VT,
which is anisotropic reflecting the anisotropy of the system.
Turning off suddenly both the applied flow and the mutual friction
transforms this VT into that of Fig. \ \ref{comp. VT}(b) after some time steps;
this VT is nearly isotropic taking $I_{\|} \simeq 0.7$; the
little deviation from the isotropic value $I_{\|}=2/3$ may be
attributed to the anisotropic boundary conditions.

The comparison of Fig. \ \ref{comp. VT}(a) and (b) shows
a marked difference.
The VT with the mutual friction consists of relatively
smooth vortex lines, while the VT without it is
very kinked owing to the lack of the smoothing effect of the
mutual friction.
Here it is necessary to check the accuracy of the numerical
calculation.
The LIA must conserve the VLD $L(t)$, whereas each numerical
procedure of reconnection can change the local line length
by a small quantity less than $\Delta \xi$ before and after the event.
We can monitor every reconnection in the VT dynamics, thus
confirming that $L(t)$ is conserved completely within the
numerical error except for at each moment of reconnection.
Then we find that our calculation is enough accurate.

\begin{figure}
\epsfxsize=8cm \epsfbox{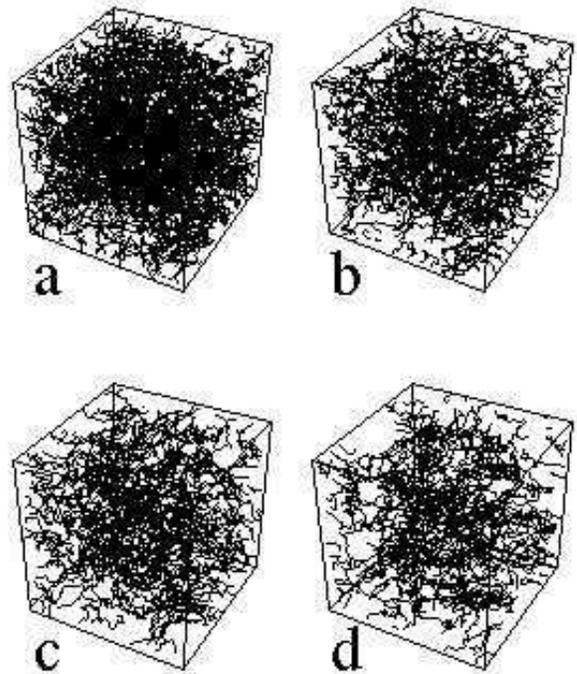}
\vspace*{3mm}
\caption{Decay of a dense VT by the LIA calculation without the mutual
friction. The system is a 1cm$^3$ cube. A periodic boundary condition is
used only along the vertical direction in these figures, while the other
sets of faces are treated as smooth, rigid walls.
The time is $t=$0s(a), 30s(b), 60s(c) and 90s(d).}
\label{decay of VT}
\end{figure}

Figure \ \ref{decay of VT} shows the decay of the VT
without mutual friction.
It is apparent that the tangle is becoming dilute.
During this process, as shown in Fig. \ \ref{decay of VT L},
$L(t)$ is actually reduced, with keeping the VT nearly
isotropic with $I_{\|} \simeq 0.7$.
Since this system is free from the mutual friction, the only
mechanism for the VT decay is that cut-off procedure which
eliminates the small vortices whose size is less than the
numerical space resolution.
However it should be noted that the continuous reduction of
$L(t)$ results in the presence of the stationary
cascade process wherein large vortices break up to smaller
ones through reconnections.
This is because, if such cascade process is absent, even
though the system is subject to that cut-off procedure,
the VT only decays a little instantaneously  and the continuous
decay is never sustained.
Only the cascade process that keeps supplying the small
vortices can reduce the VT constantly.

\begin{figure}
\epsfxsize=8cm \epsfbox{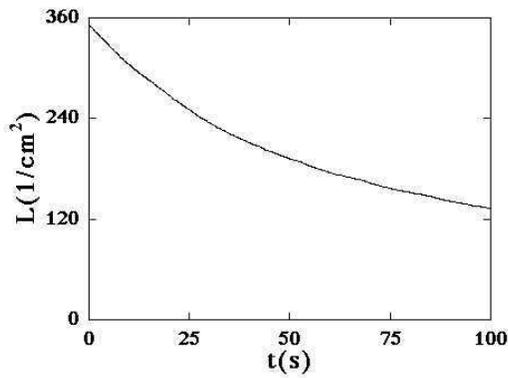}
\vspace*{3mm}
\caption{Comparison of the VLD decay for the different space resolutions
$\Delta \xi = 1.83 \times 10^{-2}$cm (solid line) and its quarter
(dotted line).}
\label{space resolution}
\end{figure}

Figure \ \ref{space resolution} compares the decay of $L(t)$
for the original space resolution $\Delta \xi$ and its quarter
$\Delta \xi/4$; the latter calculation is made by the finer time
resolution $\Delta t/16$.
The decay rate is found to be almost independent of the space resolution.
Although more coarse space resolution would affect the decay rate,
ours turn out to be enough fine to describe the cascade process.

What does this independence of the space resolution mean?
If the original resolution $\Delta \xi$ is improved to its quarter,
the vortices of the size from $\Delta \xi$ and to $\Delta \xi/4$,
which are supposed to vanish for the resolution $\Delta \xi$,
should still survive for the renewed one $\Delta \xi/4$.
Investigating the size distribution of vortices shows that
the line length of the vortices of the size between $\Delta \xi$
and $\Delta \xi/4$ is not negligible compared with the total line
length.
Nevertheless the decay of $L(t)$ little depends on the space
resolution, which is understood by the dynamical scaling
described in Sec. II.
A small vortex whose size is reduced by a factor $\lambda$ follows
the dynamics whose time scale is shortened by $\lambda ^2$.
Accordingly the small surviving vortices between $\Delta \xi$
and $\Delta \xi/4$ follow the rapid cascade dynamics to reach
the cut-off scale $\Delta \xi/4$, which proceeds much faster
than the overall decay of $L(t)$ that includes the slow dynamics
of large vortices too.
Since it is difficult to improve the space resolution furthermore
because of the computational constraints, we made the cut-off scale
coarse oppositely keeping the spare resolution $\Delta \xi$,
in order to check how the decay rate is affected.
When the cut-off scale is increased to$2\Delta \xi$, $3\Delta \xi$
and $4\Delta \xi$, the decay rate of $L(t)$ is found to be almost
the same as that with the cut-off scale $\Delta \xi$, though more
reduction of small vortices leads to larger fluctuation of $L(t)$.
Accordingly, the decay rate is independent of the space resolution
and the cut-off scale as far as we investigate in this work.
This means that the overall decay rate of the VLD is determined
principally by not small vortices but large ones whose size is
comparable to the average line spacing.

It is important to know how this behavior depends on the scale of the
system.
Section II describes that the vortex dynamics under the LIA is subject
to the dynamical scaling.
Exactly speaking, this dynamical scaling is approximate, because
the logarithmic term that depends on the characteristic radius
$<R>$ through $\beta$ is neglected(Eq. (\ref{beta})).
The logarithmic dependence is so weak that the dynamical scaling
is expected to be realized well, which should be confirmed numerically.
We made the calculation for the systems with the different scaling
factors $\lambda = 1, 10^{-1}, 10^{-2}$.
The dynamical scaling states that the VLD satisfies the relation
$L(\lambda) = \lambda^{-2} L(\lambda=1)$ \upcite{Schwarz88},
which was found to be well realized in the decay of the VT.
Hence the VT dynamics is subject to the dynamical scaling
within very high accuracy, thus being considered to be self-similar.

\begin{figure}
\epsfxsize=8cm \epsfbox{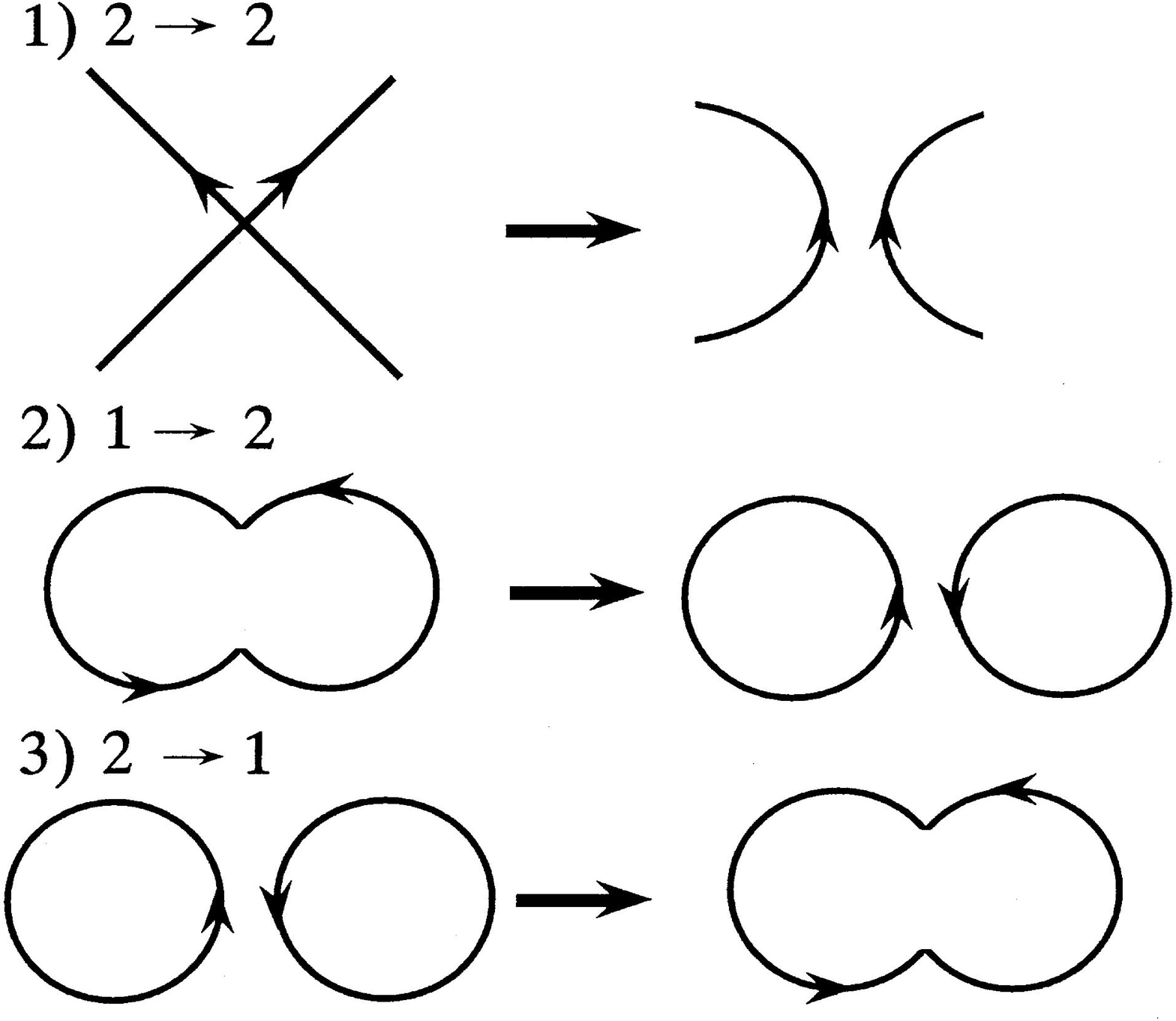}
\vspace*{3mm}
\caption{Kinds of reconnection. The type (1) shows two vortices reconnect
to two. The type (2) is one vortex divides to two (the split type).
The type (3) is two vortices are combined to one (the combination type).}
\label{topo reco}
\end{figure}

It is possible to classify the kinds of reconnection in the VT
dynamics.
The vortex reconnection is divided topologically into three classes,
as shown in Fig. \ \ref{topo reco}.
The first refers to the process whereby two vortices reconnect to
two vortices, which is most usual.
The second is the process which divides one vortex into two vortices
(the split type); the cascade process is driven by this kind of
reconnection.
Third is the process whereby two vortices are combined to one vortex
against the cascade process (the combination type).
Table 1 shows the number of reconnection events for each period in
the VT dynamics of Fig. \ \ref{decay of VT}.
The column "total" refers to the total event number of all reconnections
\upcite{comment}, and the columns "split" and "comb." represent
the event number of the above split and combination type, respectively.
Most of reconnections belong to the first class.
The reconnection of the second split type occupies about 17$\%$ of the
total reconnections, being superior to that of the third combination type of
about 10$\%$.
It is found that the reconnection of the split type actually promotes
the cascade process, against the reverse process due to that of the
combination type.

\begin{figure}
\epsfxsize=8cm \epsfbox{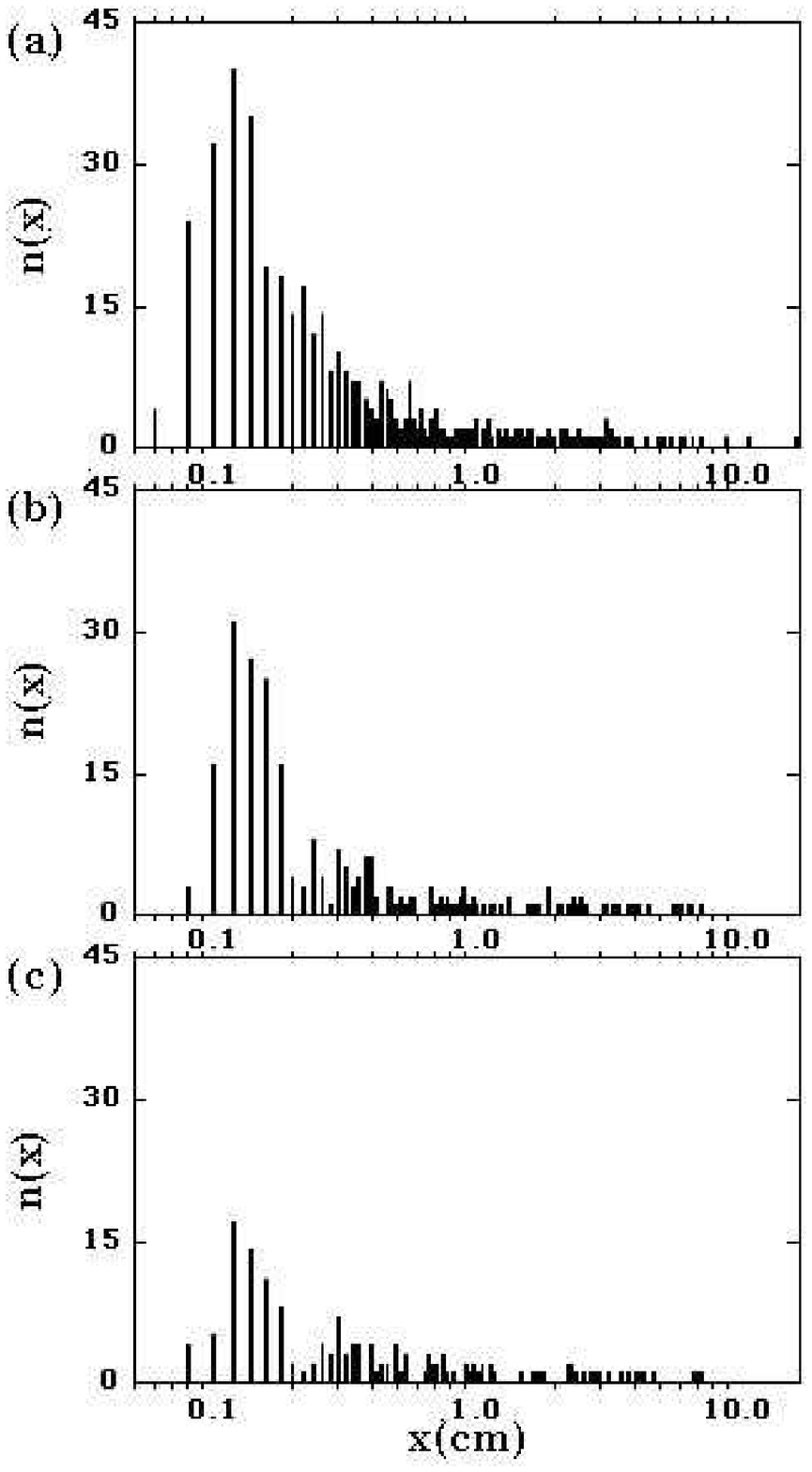}
\vspace*{3mm}
\caption{Bar chart showing the number of vortices $n(x)$ as a function
of the length $x$ in the dynamics of Fig.
\ \ref{decay of VT}.
The range of $x$ is discretized by each $\Delta x=2 \times 10^{-2}$cm.
The time is $t=$0s(a), 50s(b) and 100s(c). }
\label{l-n}
\end{figure}

\begin{figure}
\epsfxsize=8cm \epsfbox{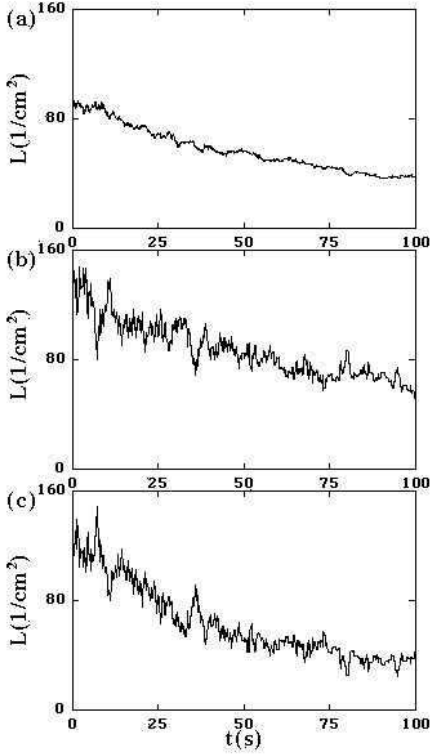}
\vspace*{3mm}
\caption{Contribution to the VLD from
the vortices in the size range (a) $[\Delta \xi, a]$,
(b) $[a, 4a]$  and (c) $[4a-]$, where $a=1$cm. }
\label{each L}
\end{figure}

The cascade process is revealed further by investigating the size
distribution of vortices.
Figure \ \ref{l-n} shows the change of the size distribution in the VT
dynamics of Fig. \ \ref{decay of VT}.
Each figure shows the number $n(x)$ of vortices as a function of
their length $x$.
The system size $a$(=1cm) and the space resolution $\Delta \xi(=1.83
\times 10^{-2}$cm), i.e., the cut-off length are the characteristic
scales in this system.
The vortices longer than $a$ are originally few,
and most vortices are concentrated in the scale range $[\Delta \xi,a]$.
As the cascade process progresses, every vortex generally divide
into smaller ones through the split type reconnections, although
some combination type reconnections may occur.
As a result, the vortices larger than $a$ become fewer, and the
vortices between $\Delta \xi$ and $a$ are decreased in number too
because they become smaller than $\Delta \xi$ and be
eliminated.\upcite{comment2}
However the contribution to the VLD is just different.
Figure \ \ref{each L} shows the contribution to the VLD from
the vortices in the size range $[\Delta \xi, a]$,
$[a, 4a]$, $[4a-]$, respectively.
The contribution from three ranges are comparable.
The VLD of the large vortices fluctuates because they are few.
The smooth VLD due to the vortices in the range  $[\Delta \xi, a]$
seems to be similar to the overall $L(t)$ of
Fig. \ \ref{decay of VT L}.
In the late stage($t \geq 50$s) of the dynamics, the large vortices
become fewer, so that the contribution of the vortices between
$\Delta \xi$ and $a$ to the overall VLD is increased relatively.

The final destiny of small vortices through the cascade process may
be interpreted several ways.
First, the vortices whose size is eventually reduced to the order
of the interatomic distance no longer sustain the vortex state,
probably changing into such short-wavelength excitation as roton
whose energy is comparable to that of the vortex.
Secondly, the vortices can vanish at a small scale by radiating phonons,
which is discussed recently by Vinen(See Sec. VI).\upcite{Vinen00}
Both mechanisms remove the small vortices from the system.
Since both mechanisms work only at a small scale, some process that
transfers energy from a large scale to smaller scales is
necessary for the decay of the VT; this is just the cascade process.
Thirdly, in a real system, the small vortices may collide
with the vessel walls as studied in Sec. III.
Since only the vortices in the bulk are observed experimentally,
the reconnection with the walls may reduce the observed VLD
effectively.

%
%%%%%%%%%%%%%%%%%%
\section{COMPARISON WITH THE VINEN'S EQUATION}

This section compares our numerical results with the solution of
the Vinen's equation to show the good agreement between them.

\begin{figure}
\epsfxsize=8cm \epsfbox{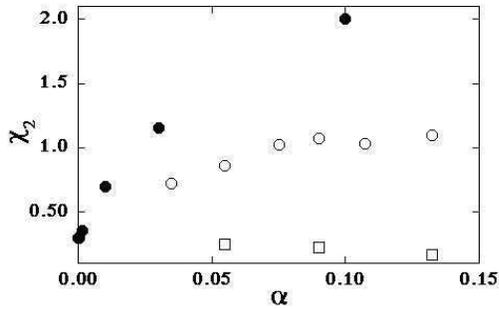}
\vspace*{3mm}
\caption{Dependence of $\chi_2$ on the mutual friction coefficient $\alpha$.
The symbols $\bullet$ show the values obtained by this work, corresponding
to $T=0$K, 0.91K, 1.07K, 1.26K, 1.6K,  in order of
increasing $\alpha$.
The symbols $\circ$ denote the values observed by Vinen\upcite{Vinen} when
a heat current is suddenly switched on, and $\Box$ the values when a heat
current is turned off.
We used the relation $\alpha = B\rho_n/2\rho$ \upcite{Donnelly} in order to
translate the Vinen's data represented by another friction coefficient $B$.}
\label{chi2}
\end{figure}

The derivation of the Vinen's equation will be reviewed
briefly.\upcite{Vinen}
Considering that cascade process at zero temperature proposed by
Feynman\upcite{Feynman}, Vinen suggested that the homogeneous
turbulence in the superflow without any normal fluid develops
in a manner analogous to that of turbulence of high Reynolds
number in an ordinary fluid.
The vortices are supposed to be approximately evenly spaced
with an average separation $\ell =L^{-1/2}$.
Then the energy of the vortices spreads from the eddies of
wave number $1/\ell$ into a wide range of wave numbers,
which means the self-similar VT sustained by the cascade process.
The overall decay of the energy density will be governed by the
chracteristic velocity $v_s=\kappa/2\pi \ell$ and the time constant
$\ell/v_s$ of the eddies of the size $\ell$, so that
\begin{equation}
\frac{d v_s^2}{d t} = -\chi_2 \frac{v_s^2}{\ell/v_s}
= -\chi_2 \frac{v_s^3}{\ell},
\end{equation}
where $\chi_2$ is a parameter.
Rewriting this by $L$, we obtain
\begin{equation}
\frac{d L}{d t}= -\chi_2 \frac{\kappa}{2\pi} L^2.
\label{Vinen1}
\end{equation}
This is the Vinen's equation that describes the decay of the VLD $L(t)$,
 and its solution is given by
\begin{equation}
\frac{1}{L} = \frac{1}{L_0} + \chi_2 \frac{\kappa}{2\pi} t,
\label{solution}
\end{equation}
where $L_0$ is the VLD at $t=0$.
At finite temperatures, the presence of the normal fluid may
affect the cascade process.
However, since the addition of the normal fluid introduces no
new dimensional parameters into the vortex dynamics, the form of
Eq.(\ref{Vinen1}) cannot be altered and $\chi_2$ becomes a function
of the temperature.
The values of $\chi_2$ observed at finite temperatures are shown
in Fig. \ \ref{chi2}.
The symbols $\circ$ denotes the values observed when a heat
current is suddenly switched on, while $\Box$ the values when
a heat current is turned off.
In any case, two kinds of $\chi_2$ reflects the complicated
behavior of the normal fluid.

\begin{figure}
\epsfxsize=8cm \epsfbox{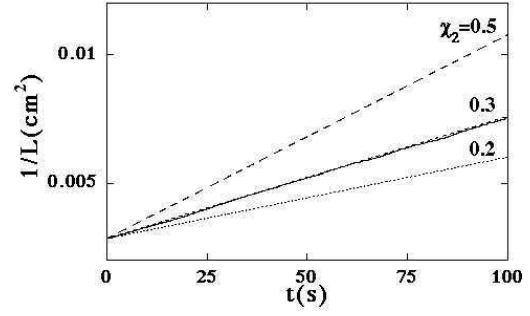}
\vspace*{3mm}
\caption{Comparison of the decay of $L(t)$ in Fig. \ \ref{decay of VT} and the
the solution (Eq. (\ref{solution})) of the Vinen's equation.
The values of $\chi_2$ as a fitting parameter for Eq. (\ref{solution}) are
shown in the figure.}
\label{comp. Vinen}
\end{figure}

Figure \ \ref{comp. Vinen} shows the comparison of our numerical
results and the solution of the Vinen's equation.
The solid line refers to our result for the VT decay of
Fig. \ \ref{decay of VT}, while three other lines denote
Eq. (\ref{solution}) with the parameters
$\chi_2=0.5,~0.3,~0.2$.
Then we find that our result agrees excellently with the solution of
$\chi_2=0.3$.
There are two meanings for this.
First, the decay of the numerical VT is well described by the Vinen's
equation.
As stated in the last paragraph, the Vinen's equation is based closely
on the cascade process.
Hence their agreement supports that the cascade process occurs really
in the numerical simulation.
Secondly, as seen from Fig. \ \ref{chi2}, the two kinds of data
$\circ$ and $\Box$ are extrapolated towards zero temperature, then
seeming to reach reasonably to $\chi_2 \simeq 0.3$; the value obtained
numerically may be consistent with those observed at finite temperatures.

In order to study how the mutual friction affects the cascade process,
we calculate the decay of the VT with the mutual friction under the
static normal fluid.
As noted by Barenghi and Samuels\upcite{BarenghiSamuels}, such phenomena
might as well be calculated not kinematically but by a self-consistent
approach which takes into account the back reaction of the VT onto the
normal fluid.
However, since the decay of an approximately isotropic and homogeneous
VT may not induce some overall flow in the static normal fluid, this work,
for simplicity, calculates kinematically the problem subject to the
static normal fluid.
Similar to the above calculation, we compare the numerical decay of the
VT at finite temperatures with Eq. (\ref{solution}) with a fitting
parameter $\chi_2$.
The obtained dependence of $\chi_2$ on the mutual friction coefficient
$\alpha$ is also shown in Fig. \ \ref{chi2}.
When the temperatures are relatively low ($T=$0.91K, 1.07K and
1.26K), the solution with a proper value of $\chi_2$ can describe
well the numerical result.
However, as the temperature increases ($T=$1.6K), the numerical
results become to deviate from Eq. (\ref{solution}).
This seems to be reasonable.
The decay term of the Vinen's equation was derived originally  based
on the idea of the homogeneous turbulence.\upcite{Vinen}
At low temperatures, the mutual friction is too small to disturb the
inertial range, while the mutual friction at high temperatures shrinks
not only small vortices but also large ones, thus disturbing the
inertial range and deviating the numerical result from
Eq. (\ref{solution}).
%
%%%%%%%%%%%%%%%%%%
\section{CONCLUSIONS AND DISCUSSIONS}

Motivated by the recent experimental work by Davis et.al.\upcite{Davis},
we studied numerically the dynamics of the VT without the mutual
friction.
The absence of the mutual friction means that the usual well-known
mechanism does not work for its free decay, so that we do not know why
the VT decays.
Throughout this paper, we conclude that the self-similar cascade
process whereby large vortex loops break up to smaller ones proceeds
in the VT, being closely concerned with the decay of the VT.
This cascade process, which may be covered with the mutual friction
at high temperatures, is just the one at zero
temperature Feynman proposed \upcite{Feynman},
although the eventual destiny of the
minimum vortex ring is beyond this formulation.
The full Biot-Savart calculation is made for dilute vortices, while
the LIA calculation for the dense VT.
The former reveals the scenario: the reconnection of the vortices
excites vortex waves on them and makes the vortex lines kinked,
which would be suppressed in the presence of the mutual friction.
The kinked parts reconnect with the body loop they belong to,
breaking up to small loops.
The LIA calculation shows that the cascade process proceeds in the
VT, keeps making the small vortices below the space resolution and
reduces the VLD $L(t)$.
Although the small vortices below the space resolution are eliminated
numerically, it should be emphasized that the VT never decays without
the cascade process.
The decay of $L(t)$ obtained numerically is consistent with the
solution of the Vinen's equation.
The calculation that takes account of the mutual friction shows
that both the modified cascade process and the vortex shrinkage
due to the mutual friction proceeds together in the VT at a finite
temperature.

\begin{figure}
\epsfxsize=8cm \epsfbox{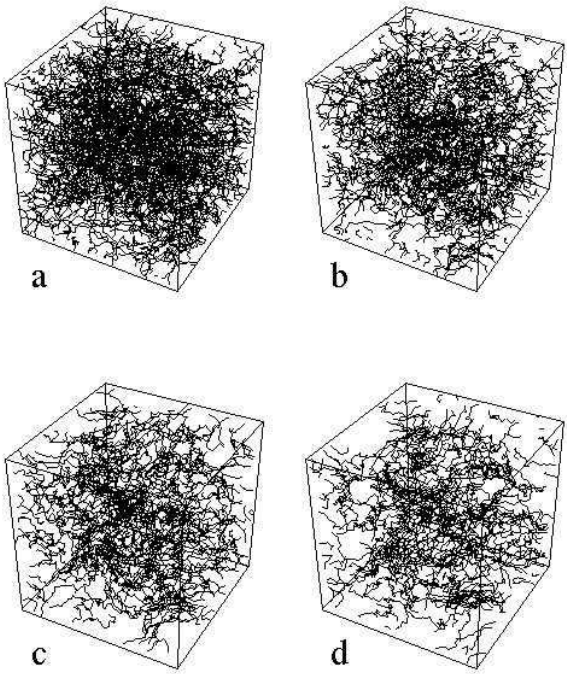}
\vspace*{3mm}
\caption{Decay of the VLD $L(t)$ for the dynamics of Fig. \ \ref{decay
of VT}.}
\label{decay of VT L}
\end{figure}

Here we will describe the recent work by Vinen.\upcite{Vinen00}
In relation to the experimental work of the grid turbulence\upcite{Stalp},
Vinen discussed the dissipation of the VT at zero temperature.
The dissipation can occur only by the emission of sound waves (phonons)
by an oscillating vortex.
The vortex oscillation of the average vortex spacing $\ell = L^{-1/2}$
has the characteristic velocity $v_{\ell} \sim \kappa/\ell$ and
the characteristic time $\tau_{\ell} \sim \ell^2/\kappa$.
Estimating the dipole and quadrupole radiation from a Kelvin wave
finds that such oscillation can cause only the very slow decay of
the VT compared with $\tau_{\ell}$.
Hence Vinen considered the excitation of the Kelvin wave whose
wavelength is much smaller than $\ell$.
In a classical viscous fluid, there is a flow of energy from components
of the velocity field with small wave numbers to components with large
wave numbers, energy being dissipated by viscosity near the Kolmogorov
wave number.
The superfluid system will have the energy cascade process of the
Kelvin waves, whereby the energy is transformed to Kelvin waves
with wave numbers greater than $\ell^{-1}$ and eventually dissipated
at a wave number $\tilde{k}_2$ by sound radiation.
Based on this picture, Vinen reformulated the Vinen's equation
 and obtained
\begin{equation}
\tilde{k}_2 \ell = \left( \frac{C\ell}{A^{1/2} \kappa}\right)^{1/2}
\label{k2}
\end{equation}
for the case of dipole radiation, where $C$ is the speed of sound and
$A$ is a constant.
It should be noted that this
Vinen's Kelvin wave cascade process corresponds to our cascade
process which is shown by the direct simulation of the vortex dynamics.
The difference is that, although Vinen considered only the Kelvin wave,
our cascade process includes not only the excitation of vortex waves
but also the breakup of large loops to
smaller ones through reconnection, which was assumed to be negligible
by Vinen but is found to be present by our simulation.
Whether the excitation of vortex waves or the breakup of vortex loops,
the structure of small wave number will be produced continuously.
We will estimate Eq. (\ref{k2}) for our simulation of the decay of the
dense VT.
As shown in Fig. \ \ref{decay of VT L}, $L$ is supposed to be 400
cm$^{-2}$, so that $\ell = L^{-1/2} = 1/20$cm.
Taking $C \simeq 2 \times 10^4$cm/s for liquid helium and $\kappa
\simeq 10^{-3}$ cm$^2$/s and assuming the unknown constant $A$
is the order of 1, Eq. (\ref{k2}) yields
$\tilde{k}_2 \ell \sim 10^3$, {\it i.e.}, $\tilde{k}_2 \sim 2
\times 10^4$ cm $^{-1}$.
Since the characteristic length $\tilde{k}_2^{-1} \sim 5
\times 10^{-5}$cm for sound radiation is enough smaller than our
numerical space resolution $\Delta \xi$, our cut-off procedure may be
considered to be used for the effect of the sound radiation,
assuming the cascade process continues self-similarly also
from $\Delta \xi$ to $\tilde{k}_2^{-1}$.

We have to comment on how the nonlocal interaction acts on the VT.\upcite{Nemi}
In a VT, the local field is usually superior to the nonlocal field.
As stated in Sec. III, however, when two vortices approach each other,
the nonlocal interaction can stretch them partly.
The full Biot-Savart calculation in Sec. III shows that in
dilute vortices the cascade process is superior to the stretch due to
the nonlocal interaction.
In a dense VT, these two processes can compete with each other; which is
superior may depend on the VLD or the size distribution of vortices.
Although the full Biot-Savart calculation for a dense VT is much CPU
expensive and difficult, we start the calculation and obtain some
preliminary results showing that the decay due to the cascade process
still proceeds.
The detail will be reported shortly.

Our results are compared with the recent experiment by Davis et.al.
\upcite{Davis}
The observed $T$-independent decay below 70mK strongly suggests
that the phonon gas plays no role, because the phonon density
falls as $T^3$ in this range, and there must be an unknown
intrinsic process in this superfluid system.
We believe that our cascade process is closely connected with
the $T$-independent decay.
Davis et.al. observed the time costant of the decay was the
order of 10 sec.
The time constant depends on the amplitude of the VLD, but we
do not know exactly the homogeneity of the VT and the amplitude
of the VLD in the experiments.\upcite{Pri}
Accordingly it is difficult to compare our results quantitatively
with the experimental data at present.

Such sound radiation can heat the fluid, which is recently discussed
by Samuels and Barenghi.\upcite{heat}
They estimated thermodynamically how much the temperature of the
fluid increases when the kinetic energy of the VT is transformed
to compressive energy, i.e., phonons.
Since the traditional second-sound technique fails in the very
low temperatures, the observation of the {\it vortex heating} is
useful for investigating this system.

Nore et.al.\upcite{Nore} studied the dynamics of the VT without any
friction, by the direct numerical simulation of the Gross-Pitaevskii
equation.
They show that the total energy of the VT is partly transformed to
compressive energy, and the energy spectrum can follow the
Kolmogorov law.
The dynamics they studied seems to include the cascade process of
this work, but its detail is not clear.

Finally we will comment on the eddy viscosity.
The superfluid turbulent state\upcite{Tough} in a capillary flow induces
excess temperature and pressure differences between both ends of the
capillary, more than those in the laminar flow state.
The excess temperature difference is understood by
the mutual friction, while the excess pressure difference is described
phenomenologically by the eddy viscosity.
The eddy viscosity works for superfluid and reduces its total momentum,
but its origin has not been necessarily revealed.
The eddy viscosity which is thought to be an intrinsic mechanism
in superfluid may be related with this cascade process.

\acknowledgments

We acknowledge W.F. Vinen and P.V.E. McClintock for useful
discussions.
One of the author(S.N.) thanks Osaka City University(OCU) for
giving an opportunity to visit OCU and Russian Foundation of Basic
Research (grant N 99-02-16942) for supporting that field.

%
%%%%%%%%%%%%%%%%%%%%

%
%%%%%%%%%%%%%%%%%%%%
\newpage
%figure captions
\newpage
\begin{table}
\caption{Classification of the reconnection events for each period
in the VT dynamics of Fig. \ \ref{decay of VT}. See the text.
\label{table1}}
\begin{tabular}{lccccr}
time(s) & total & split & comb.\tablenote{combination.}&
split/total($\%$) & comb./total($\%$) \\
\tableline
0-10 & 1921 & 298 & 203 & 15.5 & 10.6 \\
10-20 & 1366 & 252 & 161 & 18.4 & 11.8 \\
20-30 & 1021 & 178 & 114 & 17.4 & 11.2 \\
30-40 & 771 & 164 & 96 & 21.3 & 12.5 \\
40-50 & 588 & 122 & 71 & 20.7 & 12.1 \\
50-60 & 508 & 101 & 56 & 19.9 & 11.0 \\
60-70 & 393 & 58 & 26 & 14.8 & 6.6 \\
70-80 & 319 & 46 & 34 & 14.4 & 10.7 \\
80-90 & 252 & 34 & 21 & 13.5 & 8.3 \\
90-100 & 255 & 39 & 13 & 15.3 & 5.1 \\
\tableline
 & 7394 & 1292 & 795 & 17.5 & 10.8 \\
\end{tabular}
\end{table}

%%%%%%%%%%%%%%%%%%%%

\begin{thebibliography}{99}
\bibitem{comment} How the total number of reconnection events
depends on the VLD $L$ shown in Fig. \ \ref{decay of VT L} is understood
easily.
The average separation between vortices is $\ell = L^{-1/2}$ and the
characteristic self-induced velocity is $v_{\ell} = \kappa/2\pi \ell$.
Since the collision time of vortices is $\tau = \ell/v_{\ell} =
2\pi/\kappa L$ and the number of vortices in a unit volume is the order of
$L/2\pi \ell$, the total number of collisions or reconnections per unit time
is the order of $(1/\tau) L/2\pi \ell= \left( \kappa/(2\pi)^2 \right) L^{5/2}$.
The $L$-dependence of the total event number in Table 1 is consistent with
this estimation.
\bibitem{comment2} If our system had no minimum scale, the number of
vortices would increase with decreasing their sizes.
However this does not mean that the contribution of such small vortices
to the VLD becomes infinite.
\end{thebibliography}
\end{document}